# On Construction of the (24, 12, 8) Golay Codes

Xiao-Hong Peng, *Member, IEEE*, and Paddy G. Farrell, *Life Fellow, IEEE*

*Abstract*—Two product array codes are used to construct the (24, 12, 8) binary Golay code through the direct sum operation. This construction provides a systematic way to find proper (8, 4, 4) linear block component codes for generating the Golay code, and it generates and extends previously existing methods that use a similar construction framework. The code constructed is simple to decode.

## I. INTRODUCTION

THE (24, 12, 8) binary block code, denoted by $C_{24}$, was originally constructed by extending the (23, 12, 7) Golay code [1], a unique 3-error correcting perfect code. Because of the optimality and attractive structure of $C_{24}$ which is *self dual* and *doubly even* [2], it has received considerable attention, leading to a large number of construction methods. Using design theory, $C_{24}$ can be formed through a 2-(11, 6, 3) design or the 5-(24, 8, 1) Steiner system [3]. All the codewords of $C_{24}$ can also be generated by ordering a set of words of length 24 over the {0, 1} alphabet lexicographically [4]. Other approaches based on constructions in a larger field include the use of a Reed-Solomon code over $\mathbf{F}_8$ [5], the *hexacode* over $\mathbf{F}_4$ [6] and the *cubic residue code* over $\mathbf{F}_4$ [7] or the Mathieu group $\mathbf{M}_{24}$ [8]. $C_{24}$ can also be constructed using component codes with shorter length and smaller dimension. This method is based on the so-called Turyn or $|a+x|b+x|a+b+x|$ construction [9], where $a, b \in C_u$, $x \in C_v$, and $C_u$ and $C_v$ are two component codes with the same length. There are three examples utilizing this construction: i) $C_u$ is the extended (7, 4, 3) cyclic Hamming code and $C_v$ is formed by reversing the codewords of $C_u$ except for the overall parity check added through the extension [9]; ii) $C_u$ is the (8, 4, 4) first-order Reed-Muller code and $C_v$ is formed by reversing the codewords of $C_u$ except for the overall parity check [10]; and iii) $C_u$ is also the (8, 4, 4) first-order Reed-Muller code and $C_v$ is a column permutation of $C_u$ [11].

In this paper, we present a construction of the (24, 12, 8) Golay code based on two array codes. In this construction, four component codes are involved: a (3, 2, 2) single-parity-check (SPC) code, a (3, 1, 3) repetition code and two (8, 4, 4) linear block codes. We have discovered that given an (8, 4, 4) code in systematic form, there exist eight other different (8, 4, 4) codes obtained either through proper row permutation on the parity submatrix of the generator matrix of the first (8, 4, 4) code, or by applying a set of construction rules. These nine (8, 4, 4) codes (the original plus the eight others) are of the same isomorphism type (with the same length, dimension and weight distribution), but represent different code subspaces. Using the given (8, 4, 4) code, together with any one of the eight (8, 4, 4) codes obtained, in all cases leads to the construction of the (24, 12, 8) Golay code. This construction systematizes and extends our previous [12] and other existing methods, including [2], [9]-[11], that apply the $|a+x|b+x|a+b+x|$ framework and use two (8, 4, 4) codes. The code constructed can be, as in the case of related constructions [2], [11], decoded with low complexity.

## II. CONSTRUCTION METHOD

### A. The Generator matrices

The two array codes concerned are both two-dimensional product codes. A product code $C$ is formed by a direct product [2] of two component codes $C_1 = (n_1, k_1, d_1)$ and $C_2 = (n_2, k_2, d_2)$. The generator matrix, $G$, of $C$ is represented in the form of a *Kronecker product* (denoted by $\otimes$) of generator matrices of its component codes, $G_1$ and $G_2$, i.e.,

$$G = G_1 \otimes G_2 = \left(g_{i,j}^{(1)} G_2\right) \text{ or } G = G_2 \otimes G_1 = \left(g_{i,j}^{(2)} G_1\right) \quad (1)$$

where $G_1 = \left(g_{i,j}^{(1)}\right)$ and $G_2 = \left(g_{i,j}^{(2)}\right)$. The resulting code is an $(n_1 n_2, k_1 k_2, d_1 d_2)$ product code, and $G$ is of size $(k_1 k_2) \times (n_1 n_2)$.

The first array code $C$ is the (24, 8, 8) product code constituted by $C = C_1 \times C_2$, where $C_1$ and $C_2$ denote an (8, 4, 4) linear systematic block code and a (3, 2, 2) single-parity-check code, respectively. The generator matrix of $C_2$ is

$$G_2 = \begin{pmatrix} 1 & 0 & 1 \\ 0 & 1 & 1 \end{pmatrix}$$

Therefore, according to (1), the generator matrix of $C$ is given by

$$G = G_2 \otimes G_1 = \begin{pmatrix} G_1 & \mathbf{0} & G_1 \\ \mathbf{0} & G_1 & G_1 \end{pmatrix} \quad (2)$$

where $G_1$ is the $4 \times 8$ generator matrix of $C_1$, and '$\mathbf{0}$' represents a $4 \times 8$ null matrix. It is noted that the (24, 8, 8) code is also used in [13] for constructing the (24, 12, 8) code. However, the construction presented in this reference leads to a nonlinear (24, 12) code, since one of the other component codes, the (8, 3) code, used in the construction is nonlinear. No direct proof is given for showing that the distance of the code is 8, although its weight distribution turns out to be that of the extended Golay code.

The second array code $C'$ is the (24, 4, 12) product code constituted by $C' = C_1' \times C_2'$, where $C_1'$ is one of the 8 other different (8, 4, 4) codes and $C_2'$ a (3, 1, 3) repetition code with the generator matrix

$$G_2' = (1\ 1\ 1).$$

Thus the generator matrix of $C'$ is given by

$$G' = G_2' \otimes G_1' = (G_1'\ G_1'\ G_1') \quad (3)$$

where $G_1'$ is the $4 \times 8$ generator matrix of $C_1'$, and obtained from $G_1$, as described later.

$C$ and $C'$ can be regarded as code subspaces of the vector space $V_{24}$. Let $\hat{C} = C + C'$ be the code subspace of length 24 spanned by $C$ and $C'$. Each codeword $\hat{\alpha}$ in $\hat{C}$ can be expressed as a sum

$$\hat{\alpha} = \alpha + \alpha' \qquad \alpha \in C \text{ and } \alpha' \in C'. \quad (4)$$

If $C$ and $C'$ are disjoint, or, in other words, all the row vectors of $G$ and $G'$ are linearly independent, the new code subspace $\hat{C}$ can be simply referred as the *direct sum* of $C$ and $C'$. Note that the *direct sum construction* or *|u|v|-construction* [2] is a special case of the general direct sum operation adopted here. In that case, the two codes involved, e.g. $C_A$ and $C_B$ (where $u \in C_A$ and $v \in C_B$), have a special structure: the non-zero elements of the codewords of $C_A$ always correspond in position to the zero elements of the codewords of $C_B$, and vice versa. This means that $C_A$ and $C_B$ are guaranteed to be disjoint.

*Lemma* 1: Code subspaces $C$ and $C'$ defined in (2) and (3) are disjoint.

*Proof*: We need to show that for any $\alpha$ and $\alpha'$ defined in (4), $\alpha + \alpha' = 0$ implies that $\alpha = 0$ and $\alpha' = 0$. Codeword $\alpha$ can be expressed as $(a, 0, a)$, $(0, b, b)$ or $(a, b, a+b)$, where $a, b \in C_1$, according (2). Also, $\alpha'$ can be expressed as $(x, x, x)$, where $x \in C_1'$, according to (3). Suppose that $\alpha + \alpha' = 0$. Then we have

$$\alpha + \alpha' = \begin{cases} (a+x, x, a+x) = 0 & \text{(i)} \\ \text{or } (x, b+x, b+x) = 0 & \text{(ii)} \\ \text{or } (a+x, b+x, a+b+x) = 0 & \text{(iii)} \end{cases}$$

From (iii), it implies that $a + x = 0$, $b + x = 0$ and $a + b + x = 0$. Thus it is easy to conclude that $a = 0$, $b = 0$ and $x = 0$, leading to $\alpha = 0$ and $\alpha' = 0$. The same result can be attained for (i) and (ii).

□

Following Lemma 1, we now can regard code $\hat{C}$ as a result of the direct sum, denoted by $\oplus$, of $C$ and $C'$, i.e.,

$$\hat{C} = C \oplus C' \quad (5)$$

with the dimension: $dim(\hat{C}) = dim(C) + dim(C') = 8 + 4 = 12$. The generator matrix of $\hat{C}$ is therefore given by

$$\hat{G} = \begin{pmatrix} G \\ G' \end{pmatrix} = \begin{pmatrix} G_1 & \mathbf{0} & G_1 \\ \mathbf{0} & G_1 & G_1 \\ G_1' & G_1' & G_1' \end{pmatrix} \quad (6)$$

We can see from (6) that $\hat{C}$ is an $(n, k) = (24, 12)$ linear code. We also notice that the codewords in $\hat{C}$ are of the form $|a + x|b + x|a + b + x|$, where $a, b \in C_1$ and $x \in C'$, which is also the case for the Turyn [9] or cubing [11] construction. The minimum distance of $\hat{C}$ dependents on the structures of $C_1$ and $C_1'$, or $G_1$ and $G_1'$. We present next a method of obtaining $G_1$ and $G_1'$ which guarantees that the minimum distance of $\hat{C}$ is 8.

Let $C_1$ be an (8, 4, 4) systematic code. The generator matrix of $C_1$ is expressed as

$$G_1 = (\mathbf{I}_4\ \mathbf{P}). \quad (7)$$

Here $I_4$ is the $4\times 4$ identity matrix and $P$ is a $4\times 4$ parity submatrix of $G_1$. The elements of $P$ are row vectors, i.e., $P = (P_1, P_2, P_3, P_4)^T$, where $P_i$ ($i=1,2,3,4$) are chosen uniquely from the set $S = \{(1110), (1101), (1011), (0111)\}$ in any order. $S$ contains all 4-tuples with Hamming weight 3, and the weight distribution of the code is $\{N(0) =1, N(4) =14, N(8) =1\}$, where $N(x)$ represents the number of the codewords of weight $x$.

$C_1'$ can be a systematic or non-systematic code. It will be shown later that for a given systematic $G_1$ there exist eight different $G_1'$, $G_{(1)}'$, $\cdots$, $G_{(8)}'$, all leading to the construction of $C_{24}$ when applying them to (6). Among them, $G_{(1)}'$, $\cdots$, $G_{(6)}'$ are systematic, while $G_{(7)}'$ and $G_{(8)}'$ are non-systematic, which will be discussed seperately in sub-sections B and C.

### B. The Permutation Criteria for Constructing $G_{(1)}', \cdots, G_{(6)}'$

When $C_1'$ is a systematic code, its generator matrix is expressed by

$$G_1' = (I_4 \ P'). \tag{8}$$

The parity submatrix of $G_1'$, $P' = (P_1', P_2', P_3', P_4')^T$, will be generated through certain row permutations, as described later in this section, of the parity submatrix $P$ of $G_1$. Note that the row permutation concerned here only takes place on the parity submatrix of $G_1$, so it will change the structure of code subspace $C_1$. Although this is equivalent to the column permutation of $P$, the choice of using row permutation will make Theorem 1 given later in this session easier to prove. The parity submatrix $P'$ as the permutation of $P$ must satisfy the criteria stated below.

Permutation Criteria:
Cri(i) $P_i' \neq P_i$ for $1 \leq i \leq 4$; and

Cri(ii) $\{P_i', P_j'\} \neq \{P_i, P_j\}$ for $1 \leq i, j \leq 4$ and $i < j$. This means that any subset of two row vectors in $P'$ is not identical to the corresponding subset in $P$, where two subsets are said to be identical if they contain the same elements regardless of the order of the elements in the subset.

Before presenting the result of our construction given in Theorem 1, we prove a useful fact based on the above criteria in Lemma 2.

*Lemma* 2: All weight-4 codewords of $C_1$ and $C_1'$ are distinct (here $C_1'$ represents only $C_{(1)}'$, $C_{(2)}'$, $\cdots$, or $C_{(6)}'$).

*Proof*: The weight-4 codewords of the (8, 4, 4) code are generated by linear combinations of either 1, 2 or 3 row vectors of the generator matrix at a time. As both $C_1$ and $C_1'$ are systematic codes, we need only to prove in this case that any of such linear combinations of $P$ is distinct from the corresponding one that involves the same subset of row vectors of $P'$. In fact, combinations involving 1 and 2 row vectors are covered by Cri(i) and Cri(ii), respectively, while combinations involving 3 row vectors follows from Cri(i) as well since, given the total number of row vectors to be 4 for both $P$ and $P'$, Cri(i) implies that $\{P_i', P_j', P_l'\} \neq \{P_i, P_j, P_l\}$ for $i, j, l = 1, 2, 3, 4$ and $i < j < l$.

□

*Theorem 1*: The code $\hat{C}$ generated by (6), with $G_1$ defined in (7) and $G_1'$ defined in (8) and satisfying the Permutation Criteria is a (24, 12, 8) linear code; i.e., is $C_{24}$.

*Proof*: The linearity of the code is obvious as $\hat{C}$ is the direct sum of $C$ and $C'$ which both are linear. With Lemma 1, we need only to show that the minimum distance of $\hat{C}$ is 8. The weights of the codewords of both $C_1$ and $C_1'$ are 0, 4 or 8. From the structure of the codewords of $\hat{C}$: $\hat{c} = (a+x, b+x, a+b+x)$ where $a, b \in C_1$ and $x \in C_1'$, it follows, by applying Lemma 2, that $wt(e+x) \geq 2$ for any $e \in C_1$ and $x \in C_1'$, where $e$ can be either $a$, $b$ or $a+b$, and $wt(y)$ represents the weight of $y$. It is obvious that $wt(e+x) \geq 2$ is a sufficient condition for $wt(\hat{c}) \geq 8$, except in the case where $a$, $b$, $a+b$ and $x$ all have weight 4 and $a \neq b$. This case has two sub-cases: if $wt(a+x) = 2$ and $wt(b+x) = 2$, then the condition $wt(a+b) = 4$ ensures that $wt(a+b+x) = 4$, because the support ( positions of non-zero elements) of $x$ must overlap with the supports of both $a$ and $b$ in two positions; if $wt(a+x) = 2$ and $wt(b+x) = 4$, or vice-versa, the supports overlap in only one position and $wt(a+b+x)$ =6. In either sub-case $wt(\hat{c}) \geq 8$.

□

It has been shown [14] that $C_{24}$ is unique in terms of the code parameters such as the length, dimension, minimum distance and weight distribution, thus the construction presented here generates the (24, 12, 8) Golay code.

Suppose that the parity submatrix of $G_1$ is given as $P = (P_1, P_2, P_3, P_4)^T$. We will show here that there exist six permutations of $P$, $P'$, that all satisfy the criteria and can be used to construct $G_1'$ and consequently $C_{24}$. Form an index set $L = \{1, 2, 3, 4\}$ representing the elements of $P$. The six permutations satisfying the criteria can be found by checking each of the 4!=24 permutations of $L$ against the criteria, as follows:

$$\begin{cases} L^{(1)} = \{3,1,4,2\}, \ L^{(2)} = \{4,1,2,3\}, \ L^{(3)} = \{2,4,1,3\}, \\ L^{(4)} = \{4,3,1,2\}, \ L^{(5)} = \{2,3,4,1\}, \ L^{(6)} = \{3,4,2,1\}. \end{cases} \quad (9)$$

Applying one-to-one mappings of these permutations onto $P$ gives six corresponding permutations, $P' = P^{(r)}$ for $1 \leq r \leq 6$, and consequently six different $G'_1$; i.e., $G'_{(1)}, \cdots, G'_{(6)}$, respectively. This will be the case for any choice of systematic $G_1$, as the number of such permutations is always given by $3! = 6$. This is because, in order to meet the criteria, the first row vector of the parity submatrix can be placed in any of 3 other positions (rows); then the vector thus displaced can be placed in any of 2 other positions (not its original position nor the original position of the first vector); in the same way, the third vector can only be placed in one designated position; and finally the fourth vector is placed in the only remaining position.

### C. The construction Rules for $G'_{(7)}$ and $G'_{(8)}$

In addition to the six systematic (8, 4, 4) code subspaces generated by $G'_{(1)}, \cdots, G'_{(6)}$, there are two more (8, 4, 4) code subspaces, which also satisfy the conditions for constructing $C_{24}$ using (6) with the given $C_1$. However, these two codes are not systematic since they have, unlike the other six codes, 8-tuples (11110000) and (00001111) as their codewords. Because of this feature, all weight-4 codewords, except the above two, of these two codes have exact two nonzero elements in each half of the 8-tuples, otherwise it will result in some 8-tuples of weight 2 or 6, e.g. (11110000) + (11101000) = (00011000) or (11110000) + (10001110) = (01111110). We call this type of weight-4 codewords *2-and-2* codewords.

Therefore, the structures of the generator matrices of these two codes, $G'_{(7)}$ and $G'_{(8)}$, will be characterized by the above features, and all weight-4 codewords generated by $G'_{(7)}$ or $G'_{(8)}$ must satisfy the statement given in Lemma 2; i.e., they are distinct from those generated by the given $G_1$. Based on these conditions, the two generator matrices are designed in the forms

$$G'_{(7)} = \begin{pmatrix} g'_{(7),1} \\ g'_{(7),2} \\ g_h \\ g_{w=8} \end{pmatrix} \text{ and } G'_{(8)} = \begin{pmatrix} g'_{(8),1} \\ g'_{(8),2} \\ g_h \\ g_{w=8} \end{pmatrix} \quad (10)$$

respectively. Here $g_h$ is an 8-tuple with all 4 nonzero elements in its either half, i.e., (00001111) or (11110000), and $g_{w=8}$ is a weight-8 8-tuple. For $g'_{(7),m}$ and $g'_{(8),m}$ ($m = 1, 2$), they are weight-4 8-tuples or 2-and-2 codewords. So when they are divided into two equal halves, i.e., left half: $g'^{(L)}_{(7),m}$ and $g'^{(L)}_{(8),m}$; and right half: $g'^{(R)}_{(7),m}$ and $g'^{(R)}_{(8),m}$, any of these halves is a weight-2 4-tuple. The constructions of $g'_{(7),m}$ and $g'_{(8),m}$ follow the rules stated below.

Construction Rules:
Rule(i) Assign
$$g'^{(L)}_{(7),1} = g'^{(L)}_{(8),1} = x \text{ and}$$
$$g'^{(L)}_{(7),2} = g'^{(L)}_{(8),2} = y,$$
where $x$ and $y$ are any two different weight-2 4-tuple with $\bar{x} \neq y$ ($\bar{v}$ is the complement of $v$).

Rule(ii) $g'^{(R)}_{(l),m}$ ($l = 7, 8$ and $m = 1, 2$) must satisfy
$$g'^{(R)}_{(7),m} \neq g'^{(R)}_{(8),m} \text{ and } \overline{g'^{(R)}_{(7),m}} \neq g'^{(R)}_{(8),m} \text{ for } m = 1, 2;$$
$$g'^{(R)}_{(l),1} \neq g'^{(R)}_{(l),2} \text{ and } \overline{g'^{(R)}_{(l),1}} \neq g'^{(R)}_{(l),2} \text{ for } l = 7, 8.$$

Rule(iii) Assume that the nonzero elements of $x$ are in its $i$-th and $j$-th positions, and that the nonzero elements of $y$ are in its $i'$-th and $j'$-th positions, where $i, j, i', j' \in \{1, 2, 3, 4\}$. The following conditions must also be satisfied.
$$g'^{(R)}_{(l),1} \neq (P_i + P_j) \text{ and } \overline{g'^{(R)}_{(l),1}} \neq (P_i + P_j) \text{ for } l = 7, 8;$$
and
$$g'^{(R)}_{(l),2} \neq (P_{i'} + P_{j'}) \text{ and } \overline{g'^{(R)}_{(l),2}} \neq (P_{i'} + P_{j'}) \text{ for } l = 7, 8.$$
Here $P_u$ ($1 \leq u \leq 4$) are the row vectors of $P$ in $G_1$ given in (7).

*Theorem 2*: Applying $G'_{(7)}$ or $G'_{(8)}$ given in (10) and satisfying the Construction Rules to the generator matrix given in (6), as $G'_1$, together with $G_1$ defined in (7), generates $C_{24}$.

The Proof of Theorem 2 is provided in the APPENDIX.

*Example:* Given $G_1$ as the generator matrix of a systematic (8, 4, 4) code, i.e.,

$$G_1 = \begin{pmatrix} 1 & 0 & 0 & 0 & 1 & 1 & 0 & 1 \\ 0 & 1 & 0 & 0 & 0 & 1 & 1 & 1 \\ 0 & 0 & 1 & 0 & 1 & 1 & 1 & 0 \\ 0 & 0 & 0 & 1 & 1 & 0 & 1 & 1 \end{pmatrix}$$

and $$P = \begin{pmatrix} (1\ 1\ 0\ 1) \\ (0\ 1\ 1\ 1) \\ (1\ 1\ 1\ 0) \\ (1\ 0\ 1\ 1) \end{pmatrix} = \begin{pmatrix} P_1 \\ P_2 \\ P_3 \\ P_4 \end{pmatrix} = (P_1, P_2, P_3, P_4)^T,$$

and using the results given in (9), the six corresponding permutations of $P$, $P^{(r)}$ ($1 \leq r \leq 6$), follow accordingly:

$$P^{(1)} = (P_3, P_1, P_4, P_2)^T = \begin{pmatrix} 1 & 1 & 1 & 0 \\ 1 & 1 & 0 & 1 \\ 1 & 0 & 1 & 1 \\ 0 & 1 & 1 & 1 \end{pmatrix},$$

$$P^{(2)} = (P_4, P_1, P_2, P_3)^T = \begin{pmatrix} 1 & 0 & 1 & 1 \\ 1 & 1 & 0 & 1 \\ 0 & 1 & 1 & 1 \\ 1 & 1 & 1 & 0 \end{pmatrix},$$

$$P^{(3)} = (P_2, P_4, P_1, P_3)^T = \begin{pmatrix} 0 & 1 & 1 & 1 \\ 1 & 0 & 1 & 1 \\ 1 & 1 & 0 & 1 \\ 1 & 1 & 1 & 0 \end{pmatrix},$$

$$P^{(4)} = (P_4, P_3, P_1, P_2)^T = \begin{pmatrix} 1 & 0 & 1 & 1 \\ 1 & 1 & 1 & 0 \\ 1 & 1 & 0 & 1 \\ 0 & 1 & 1 & 1 \end{pmatrix},$$

$$P^{(5)} = (P_2, P_3, P_4, P_1)^T = \begin{pmatrix} 0 & 1 & 1 & 1 \\ 1 & 1 & 1 & 0 \\ 1 & 0 & 1 & 1 \\ 1 & 1 & 0 & 1 \end{pmatrix},$$

and $P^{(6)} = (P_3, P_4, P_2, P_1)^T = \begin{pmatrix} 1 & 1 & 1 & 0 \\ 1 & 0 & 1 & 1 \\ 0 & 1 & 1 & 1 \\ 1 & 1 & 0 & 1 \end{pmatrix}.$

Generator matrices $G'_{(1)}, \cdots, G'_{(6)}$ will be composed of $I_4$ and one of the $P^{(r)}$ listed above. For $G'_{(7)}$ or $G'_{(8)}$, we follow Rule(i) by choosing $x = (0\,1\,0\,1)$ for $g'^{(L)}_{(7),1}$ and $g'^{(L)}_{(8),1}$; and $y = (0\,0\,1\,1)$ for $g'^{(L)}_{(7),2}$ and $g'^{(L)}_{(8),2}$. We then follow Rule(ii) and Rule(iii) by assigning: $g'^{(R)}_{(7),1} = (0\,1\,1\,0)$, $g'^{(R)}_{(8),1} = (0\,1\,0\,1)$; and $g'^{(R)}_{(7),2} = (0\,0\,1\,1)$, $g'^{(R)}_{(8),2} = (1\,0\,0\,1)$. Finally, we have

$$G'_{(7)} = \begin{pmatrix} g'_{(7),1} \\ g'_{(7),2} \\ g_h \\ g_{w=8} \end{pmatrix} = \begin{pmatrix} g'^{(L)}_{(7),1} & g'^{(R)}_{(7),1} \\ g'^{(L)}_{(7),2} & g'^{(R)}_{(7),2} \\ g_h \\ g_{w=8} \end{pmatrix} = \begin{pmatrix} 0 & 1 & 0 & 1 & 0 & 1 & 1 & 0 \\ 0 & 0 & 1 & 1 & 0 & 0 & 1 & 1 \\ 0 & 0 & 0 & 0 & 1 & 1 & 1 & 1 \\ 1 & 1 & 1 & 1 & 1 & 1 & 1 & 1 \end{pmatrix}$$

and $$G'_{(8)} = \begin{pmatrix} g'_{(8),1} \\ g'_{(8),2} \\ g_h \\ g_{w=8} \end{pmatrix} = \begin{pmatrix} g'^{(L)}_{(8),1} & g'^{(R)}_{(8),1} \\ g'^{(L)}_{(8),2} & g'^{(R)}_{(8),2} \\ g_h \\ g_{w=8} \end{pmatrix} = \begin{pmatrix} 0 & 1 & 0 & 1 & 0 & 1 & 0 & 1 \\ 0 & 0 & 1 & 1 & 1 & 0 & 0 & 1 \\ 0 & 0 & 0 & 0 & 1 & 1 & 1 & 1 \\ 1 & 1 & 1 & 1 & 1 & 1 & 1 & 1 \end{pmatrix}.$$

$C_{24}$ can then be constructed by using $G_1$ and $G'_1$ in the framework of $\hat{G}$ given in (6), where $G'_1$ can be any one of the $G'_{(1)}, \cdots, G'_{(8)}$ formed above. The codes generated using different $G'_1$ will have the same weight distribution: $\{N(0) = 1, N(8) = 759, N(12) = 2576, N(16) = 759, N(24) = 1\}$.

### III. REMARKS

The example given above shows that for the given generator matrix of a systematic (8, 4, 4) code, $G_1 = (I_4 \; P)$, there exist eight different $G'_1$, $G'_{(r)} = (I_4 \; P^{(r)})$ ($1 \le r \le 6$) plus $G'_{(7)}$ and $G'_{(8)}$, that all lead to the construction of $C_{24}$. These eight different $G'_1$ and the original $G_1$ represent different (8, 4, 4) code subspaces, although they have the same code parameters

$(n, k, d)$ and weight distribution $\{N(0) = 1, N(4) = 14, N(8) = 1\}$. We have observed some interesting properties among this group of subspaces. Denoted by $S^{(i)}_{w=4}$ ($0 \le i \le 8$) the set of weight-4 codewords generated by an (8, 4, 4) linear code over $GF(2)$, and $S^{(0)}_{w=4}$ is generated by $C_1$ and $S^{(1)}_{w=4}, \cdots, S^{(8)}_{w=4}$ by $C'_{(1)}, \cdots, C'_{(8)}$, respectively. Thus the following properties exist:

Pro(i) $|S^{(i)}_{w=4}| = 14$ for $0 \le i \le 8$.

Pro(ii) $S^{(0)}_{w=4} \cap S^{(i)}_{w=4} = \{\;\}$ for $1 \le i \le 8$.

Pro(iii) $|S^{(i)}_{w=4} \cap S^{(j)}_{w=4}| = 2$, $i, j \in \{1, \ldots, 8\}$ and $i \ne j$.

Pro(i) is obvious as the weight-4 codewords are the results of the linear combinations of either 1, 2 or 3 row vectors of the generator matrix of the (8, 4, 4) code, i.e., $|S^{(i)}_{w=4}| = \sum_{j=1}^{3} \binom{4}{j} = 14$. Pro(ii) is guaranteed by Lemma 2 for $G'_{(1)}, \cdots, G'_{(6)}$ and by the Construction Rules and (10) for $G'_{(7)}$ and $G'_{(8)}$. Pro(iii) is also true based on the Permutation Criteria and the Construction Rules. If one wishes to generate some weight-4 codewords that belong to more than two codes of $\{C'_{(1)}, \cdots, C'_{(8)}\}$, then one or more permutation pattern(s) $L^{(r)}$ ($1 \le r \le 6$) of (9) need(s) to be replaced, which will cause the Permutation Criteria to be unsatisfied. The same consequence applies to the Construction Rules if similar changes are required for $g'_{(7),m}$ and $g'_{(8),m}$ ($m = 1, 2$).

According to Pro(iii), the number of the distinct weight-4 codewords contributed by the eight codes $C'_{(1)}, \cdots, C'_{(8)}$ is given by

$$\left| \bigcup_{i=1}^{8} S^{(i)}_{w=4} \right| = 14 + (14-2) + (14-4) + \ldots + (14-12)$$
$$= \sum_{j=0}^{6} (14 - 2j)$$
$$= 56$$

Furthermore, the total number of the distinct weight-4 codewords from the nine codes, $C_1$ and $C'_{(1)}, \cdots, C'_{(8)}$ is

$$\left| \bigcup_{i=0}^{8} S^{(i)}_{w=4} \right| = |S^{(0)}_{w=4}| + 56$$
$$= 14 + 56$$
$$= 70$$

This figure means that the nine (8, 4, 4) code subspaces used in our scheme for the purpose of constructing $C_{24}$ have contained all the possible weight-4 8-tuples, since $\binom{8}{4} = 70$. This property is demonstrated in TABLE I, where all the possible weight-4 codewords generated by $G_1$ and $G'_{(1)}, \cdots, G'_{(8)}$ in the above example are listed as integers which are the decimal values of the corresponding binary 8-tuple. Integer 29, for instance, is the decimal value of the 8-tuple or codeword (10111000), calculated by taking the high-order digits on the right. The 70 distinct weight-4 codewords are indicated in bold. Also in TABLE I, the three properties: Pro(i) - Pro(iii) among the nine code subspaces: $C_1$ and $C'_{(1)}, \cdots, C'_{(8)}$ can be clearly seen. In addition, the intersections between any pair of $\{ C'_{(1)}, \cdots, C'_{(8)} \}$ are two complementary codewords, and their corresponding integers add up to 255.

TABLE I
THE WEIGHT-4 CODEWORDS OF THE (8, 4, 4) CODES (results from *EXAMPLE*)

| Code | Weight-4 codewords (in decimal) | | | | | | | | | | | | | |
|---|---|---|---|---|---|---|---|---|---|---|---|---|---|---|
| $C_1$ | **29** | **39** | **58** | **78** | **83** | **105** | **116** | **139** | **150** | **172** | **177** | **197** | **216** | **226** |
| $C'_{(1)}$ | **23** | **43** | **60** | **77** | **90** | **102** | **113** | **142** | **153** | **165** | **178** | **195** | **212** | **232** |
| $C'_{(2)}$ | **27** | **46** | **53** | 77 | **86** | **99** | **120** | **135** | **156** | **169** | 178 | **202** | **209** | **228** |
| $C'_{(3)}$ | **30** | **45** | **51** | **75** | **85** | 102 | 120 | 135 | 153 | **170** | **180** | **204** | **210** | **225** |
| $C'_{(4)}$ | 23 | 46 | **57** | 75 | **92** | **101** | **114** | **141** | **154** | **163** | 180 | **198** | 209 | 232 |
| $C'_{(5)}$ | 30 | 43 | 53 | **71** | **89** | **108** | 114 | 141 | **147** | **166** | **184** | 202 | 212 | 225 |
| $C'_{(6)}$ | 27 | 45 | **54** | 71 | 92 | **106** | 113 | 142 | **149** | 163 | 184 | **201** | 210 | 228 |
| $C'_{(7)}$ | **15** | 51 | 60 | 86 | 89 | 101 | 106 | 149 | 154 | 166 | 169 | 195 | 204 | **240** |
| $C'_{(8)}$ | 15 | 54 | 57 | 85 | 90 | 99 | 108 | 147 | 156 | 165 | 170 | 198 | 201 | 240 |

Pro(i) - Pro(iii) also suggest a balanced incomplete block design (BIBD), (or, in some cases, the dual of a BIBD); i.e., the $(v, b, r', k', \lambda)$-configuration [15], which can be described by the incidence matrix $Q$. The rows of the matrix correspond to the sets $S^{(1)}_{w=4}, \cdots, S^{(8)}_{w=4}$, and the columns of the matrix correspond to the 56 weight-4 codewords, $c^1_{w=4}, \ldots, c^{56}_{w=4}$. The entry in the $i$-th row and the $j$-th column of $Q$ is a 1 if $S^{(i)}_{w=4}$ contains $c^j_{w=4}$ and is a 0 otherwise. Thus in $Q$, there are $v = 8$ rows and $b = 56$ columns; each row has exactly $r'= 14$ 1's; each column has exactly $k'= 2$ '1's; and the inner product of any two rows is equal to $\lambda = 2$. The $Q$ matrix of the above example is give below.

$$Q = \begin{pmatrix} 01001000000100100010001001000010010001000010010010000010010 \\ 00100010100000101000100000011000000100010100000101000100 \\ 00010101000001010000001000011000010000010100000101010000 \\ 01000010001001000001010000100100001010000100100100010000010 \\ 00011000100010000100000010101001010000001000010000110000 \\ 00101000100100000010001010000101000010000010100010000100 \\ 10000001000100001100010100000000101000110000100010000001 \\ 10000000011000010010100010000001000101001000011000000001 \end{pmatrix}$$

The relationships between the ordinal numbers of the columns of $Q$ and the 56 weight-4 codewords belonging to $S^{(1)}_{w=4}, \cdots, S^{(8)}_{w=4}$ can be easily worked out from TABLE I.

Based on the above results, our construction method presented here can be considered to have extended the existing schemes that apply the $|a + x|b + x|a + b + x|$ framework and use two (8, 4, 4) codes. For instance, the original Turyn construction employs two (8, 4, 4) codes: the extended (7, 4, 3) cyclic Hamming code and its specified version [9, p.588] generated by

$$G_{ex(7,4,3)} = \begin{pmatrix} 0 & 0 & 0 & 1 & 1 & 0 & 1 & 1 \\ 0 & 0 & 1 & 1 & 0 & 1 & 0 & 1 \\ 0 & 1 & 1 & 0 & 1 & 0 & 0 & 1 \\ 1 & 1 & 0 & 1 & 0 & 0 & 0 & 1 \end{pmatrix}$$

and $$G'_{ex(7,4,3)} = \begin{pmatrix} 1 & 0 & 1 & 1 & 0 & 0 & 0 & 1 \\ 0 & 1 & 0 & 1 & 1 & 0 & 0 & 1 \\ 0 & 0 & 1 & 0 & 1 & 1 & 0 & 1 \\ 0 & 0 & 0 & 1 & 0 & 1 & 1 & 1 \end{pmatrix}$$

respectively. This construction is just one of the 8 possible cases given in the above example, i.e., $G_1$ and $G'_1 = G'_{(4)} = (I_4 \; P^{(4)})$ obtained in *Example* are the results of applying elementary row operations over $G_{ex(7,4,3)}$ and $G'_{ex(7,4,3)}$, respectively. In the same way, it can be shown that the Forney's construction [11] employing the (8, 4, 4) first-order Reed-Muller code and its permutation is also covered by our scheme, as the choices of

$$G_1 = \begin{pmatrix} 1 & 0 & 0 & 0 & 0 & 1 & 1 & 1 \\ 0 & 1 & 0 & 0 & 1 & 1 & 0 & 1 \\ 0 & 0 & 1 & 0 & 1 & 1 & 1 & 0 \\ 0 & 0 & 0 & 1 & 1 & 0 & 1 & 1 \end{pmatrix}$$

and one of the corresponding 8 possible $G'_1$,

$$G'_{(7)} = \begin{pmatrix} 1 & 0 & 1 & 0 & 1 & 0 & 1 & 0 \\ 1 & 1 & 0 & 0 & 1 & 1 & 0 & 0 \\ 1 & 1 & 1 & 1 & 0 & 0 & 0 & 0 \\ 1 & 1 & 1 & 1 & 1 & 1 & 1 & 1 \end{pmatrix},$$

represent exactly the same code subspaces as the two generator matrices $G^*_{(8,4)}$ and $G_{(8,4)}$ given in [11, p.1174] do, respectively.

## IV. CONCLUSION

In summary, we have shown that the (24, 12, 8) binary Golay code can be constructed as the direct sum of two array codes involving four component codes, two of which are simple linear block codes (a repetition code and an SPC code). The other two component codes are two different (8, 4, 4) codes; one of them is a systematic code and the other is its modified version. There exist eight different such modified codes which meet the construction criteria or rules presented. It is not difficult to show that the eight corresponding (24, 12, 8) codes formed using our construction represent eight different, though partly overlapped, code subspaces, but they are all equivalent or isomorphic to the (24, 12, 8) Golay code.

We have also demonstrated that the generator matrix $G_1$ and eight corresponding generator matrices $G'_{(1)}, \cdots, G'_{(8)}$ involved in our construction can generate all possible codewords (including weight-4 and weight-8) that any (8, 4, 4) code can do. In this sense, the method presented here may be viewed as the generalization of all existing approaches to constructing the (24, 12, 8) Golay code using the $|a+x|b+x|a+b+x|$ construction and two (8, 4, 4) codes. It is worth pointing out that in our method the generator matrices $G'_{(1)}, \cdots, G'_{(8)}$ can also be described as the result of column permutations of $G_1$, though a lengthier proof may be required.

There are various ways to decode the (24, 12, 8) Golay code, such as the decoders based on the hexacode constructions for hard-decision [16-18] and soft-decision [19, 20] decoding. For the construction based on (6), a regular trellis can be built using different techniques [11, 21]. This trellis has three sections of length 8 and 64 states at each section boundary. Essentially, it consists of eight structurally identical sub-trellises, thus enabling simple and fast maximum likelihood decoding as these sub-trellises can be processed in parallel. Trellises with such a structure are desirable because this can considerably reduce interconnections within the IC, and the chip-size, which are major concerns in implementing trellis decoding using DSP and VLSI technologies [22, 23].

The decoding complexity can be further reduced by simplifying the sub-trellises through initial-decoding on the component code $C_1$ [24].

## APPENDIX
## PROOF OF THEOREM 2

*Theorem 2*: Applying $G'_{(7)}$ or $G'_{(8)}$ given in (10) and satisfying the Construction Rules to the generator matrix given in (6), as $G'_1$, together with $G_1$ defined in (7), generates $C_{24}$.

*Proof*: We first show that the two codes generated using $G'_{(7)}$ and $G'_{(8)}$, respectively, which are constructed base on (10) and the Construction Rules, are (8, 4, 4) codes. This is equivalent to showing that both codes can generate 14 weight-4 codewords, given the weight distribution of the (8, 4, 4) code: $\{N(0) =1, N(4) =14, N(8) =1\}$. It is apparent that the 4 rows of either $G'_{(7)}$ or $G'_{(8)}$ are linearly independent. Since $g'_{(l),1}$, $g'_{(l),2}$ and $g'_{(l),1} + g'_{(l),2}$ ($l = 7, 8$) are all weight-4 8-tuples or 2-and-2 codewords according to Rule(i) & Rule(ii), linear combinations of any 2, 3 and 4 row vectors of $G'_{(7)}$ or $G'_{(8)}$ all result in weight-4 codewords. The total number of the weight-4 codewords that $G'_{(7)}$ or $G'_{(8)}$ can generate is $3 + \sum_{j=2}^{4} \binom{4}{j} = 14$; here '3' represents the number of the rows with 4 nonzero elements of $G'_{(7)}$ or $G'_{(8)}$ except the all-one row $g_{w=8}$. Of the 14 weight-4 codewords, in addition to codewords (00001111) and (11110000), there are 12 2-and-2 codewords, owing to the special structures of $G'_{(7)}$ and $G'_{(8)}$ in (10) that contain $g_{w=8}$ and $g_h$.

We then show that all weight-4 codewords of $C_1$ and $C'_{(7)}$ or $C'_{(8)}$ are distinct. As $C_1$ is a systematic (8, 4, 4) code and $G_1$ contains a 4×4 identity matrix, it does not have the codewords (00001111) and (11110000), but has 6 codewords 2-and-2 codewords because there are in total $\binom{4}{2} = 6$ linear combinations involving 2 row vectors of $G_1$. Based on (7), these 6 codewords can be expressed as

$cw1=(1100~P_{1,2})$, $cw2=(1010~P_{1,3})$, $cw3=(1001~P_{1,4})$,
$cw4=(0011~P_{3,4})$, $cw5=(0101~P_{2,4})$, $cw6=(0110~P_{2,3})$.

Here $P_{i,j} = P_i + P_j$ for $1 \le i, j \le 4$ and $P_i, P_j \in \boldsymbol{P}$ given in (7). Obviously, the left halves of the 6 codewords are unique weight-2 4-tuples, so are the right halves otherwise some combinations of such codewords could result in the codewords of weight < 4. It is easy to show that $P_{3,4} = \overline{P_{1,2}}$, $P_{2,4} = \overline{P_{1,3}}$ and $P_{2,3} = \overline{P_{1,4}}$ since $G_1$ has the all-one codeword $g_{w=8}$. This means that these 6 codewords form 3 complementary pairs, i.e., $cw4 = \overline{cw1}$, $cw5 = \overline{cw2}$ and $cw6 = \overline{cw3}$.

The 12 2-and-2 codewords of $C'_{(7)}$ or $C'_{(8)}$ can be divided into three groups:

(i) $a, a', \overline{a}, \overline{a}'$;

(ii) $b, b', \overline{b}, \overline{b}'$; and

(iii) $a+b, (a+b)', \overline{a+b}, \overline{(a+b)'}$.

Here $a$ and $b$ represent $\mathbf{g}'_{(l),1}$ and $\mathbf{g}'_{(l),2}$ for $l = 7, 8$, respectively, and $u' = u + (00001111)$ where $u \in U = \{a, b, (a+b)\}$ and $\overline{v} = v + (11111111)$ where $v \in V = U \cup \{a', b', (a+b)'\}$. Without loss of generality, we assign $a^{(L)}$ (or $\mathbf{g}'^{(L)}_{(l),1}$) = (1100) (the left half of $cw1$) and $b^{(L)}$ (or $\mathbf{g}'^{(L)}_{(l),2}$) = (1010) (the left half of $cw2$) for $l = 7, 8$, according to Rule(i). Rule(iii) ensures that $a^{(R)} \neq P_{1,2}$, $a'^{(R)} \neq P_{1,2}$, $b^{(R)} \neq P_{1,3}$ and $b'^{(R)} \neq P_{1,3}$.

Now it can be concluded that $a$, $a'$, $b$, and $b'$ are distinct from all the codewords of $C_1$ ($cw1, \cdots, cw6$) as any corresponding pair of them differ in either left or right half of the 8-tuple. The same conclusion applies to $\overline{a}, \overline{a}', \overline{b}$ and $\overline{b}'$, since $s \neq t$ implies $\overline{s} \neq \overline{t}$. For the last group of the 12 codewords, due to the fact that $(a+b)^{(L)} = a^{(L)} + b^{(L)}$ =(1100)+(1010)=(0110) (the left half of $cw6$) and $P_{1,2} + P_{1,3} = P_{2,3}$, it can also be concluded that the 4 codewords $a+b, (a+b)', \overline{a+b}$ and $\overline{(a+b)'}$ are distinct from $cw1, \cdots, cw6$ by applying the method and properties described above.

By now we have shown that the structures of $G'_{(7)}$ and $G'_{(8)}$ given in (10) and the Construction Rules guarantee the same conclusion presented in Lemma 2; i.e., all weight-4 codewords of $C_1$ and $C'_{(7)}$ or $C'_{(8)}$ are distinct. The rest of the proof is analogous to that for Theorem 1.

□

ACKNOWLEDGMENT

The authors wish to thank the anonymous reviewers and the Associate Editor, Robert McEliece, for their detailed and helpful comments and suggestions.